

\magnification=1200
\font\large=cmbx10 at 14.4pt
\centerline{\large Tolerance and Sensitivity in the Fuse Network}
\vskip 3truecm
\centerline{G.\ George Batrouni,\footnote{$^*$}{Permanent Address:
Thinking Machines Corporation, 245 First Street, Cambridge, MA 02142, USA}
Alex Hansen\footnote{$^\dagger$}{Also at Institutt for fysikk, NTH,
N--7034 Trondheim, Norway} and Gerald H.\ Ristow}
\vskip5pt
\centerline{Groupe Mati\`ere Condens\'ee et Mat\'eriaux, URA CNRS 804}
\centerline{Universit\'e de Rennes I}
\centerline{F--35042 Rennes Cedex, France}
\vskip3.0truecm
\baselineskip=18truept plus 1 pt minus 0pt \parskip=9truept
{\noindent We show that depending on the disorder, a small noise added to
the threshold distribution of the fuse network may or may not completely
change the subsequent breakdown process.  When the threshold distribution
has a lower cutoff at a finite value and a power law dependence towards
large thresholds with an exponent which is less than $0.16\pm0.03$, the
network is not sensitive to the added noise, otherwise it is.  The
transition between sensitivity or not appears to be second order, and is
related to a localization-delocalization transition earlier observed in
such systems.}
\vskip 5cm
\noindent Submitted to Journal of Physics A
\hfill Version of September 25, 1993
\vfill
\eject
Suppose one manufactures a set of machine parts, say, which are identical
to within some predetermined tolerance.  One may ask whether this
predetermined tolerance is small enough so that one may with reasonable
certainty predict the strength and fracture properties of each member of
the set.  If it is, then testing one member of the set will give a
representative idea of what is to be expected from the other members.
However, fracture is a highly correlated process where singularities in the
stress field are caused by fractures opening, and these singularities in
turn produce more cracks. Thus, it is intuitively very likely that the
eventual fractures forming will be very sensitive to what may appear as
small initial differences between the various samples, with the result
for example that the fracture toughness varies considerably from sample to
sample.  Is it therefore possible to define concept of tolerance in the sense
that if two members
of a set is equal to within such and such limit, they will have the same
fracture properties?  It is the
aim of this letter to discuss this question. We use the fuse model,
originally introduced by de Arcangelis {\it et al.,\/}$^1$ as a model
system.  This model has proven itself to be extremely rich in addition to
capturing some of the essential features of brittle fracture --- see Ref.\
2 for a throrough discussion of this. We find that whether rupture
develops in a manner which is unpredictable in the sense discussed above
does not depend on the noise distribution, but on the distribution of local
strengths of the system itself.  For some strength distributions, the
network is sensitive to the initial added noise, and for other
distributions it is not.  We also find numerically that there is a second
order phase transition separating the sensitive from the nonsensitive
regime. We suggest that this phase transition reflects a
localization-delocalization transition previously seen in this
system.$^{3,4}$

We work with a square lattice of size $L\times L$ oriented at $45^\circ$
between two bus bars.  In the direction parallel to the bus bars, the lattice
is periodic.  Each bond in the lattice is a fuse, {\it i.e.\/} it acts as
an ohmic resistor as long as the current it carries $i$ is lower than some
threshold current $t$.  If the current exceeds this threshold, the fuse
``blows" and turns irreversibly into an insulator.  Each bond, i.e. fuse,
is assigned a threshold $t$ from a (cumulative) statistical distribution
$P(t)$.  There
are no spatial correlations build into the way the threshold values are
assigned.  After the thresholds have been assigned, we imagine setting up
a potential difference between the two bus bars, which is slowly increased.
As fuse after fuse reaches its threshold current and burns out, the
conductivity across the network decreases until it drops to zero.  At this
point a band of blown fuses has formed which cuts the network in two.

The breakdown process is highly complex, with long-range correlations
developing as it advances towards rupture.  This is most easily seen through
the way we actually simulate the breakdown process numerically.  Each time
a fuse has blown, we recalculate the current distribution within the network
by solving the Kirchoff equations for a unit voltage difference between the
bus bars.  For each bond $k$, we calculate the ratio $i_k/t_k$.  We then
search for the maximum such ratio, $\max_k (i_k/t_k)$.  The corresponding
bond is the next to be cut, and this will happen at a voltage difference
$1/\max_k (i_k/t_k)$.  In the beginning of the breakdown process when few
bonds have been cut, the current distribution is very narrow, {\it i.e.\/}
the bonds
all carry nearly the same current (and when no bonds have broken they all
carry exactly the same current).  Thus, the bonds which are likely to break,
{\it i.e.\/} those bonds whose ratio $i_k/t_k$ is large, are those whose
thresholds $t_k$ are small --- the weak bonds.  However, as the rupture
process evolves, the current distribution becomes wider and wider, and
eventually a large ratio $i_k/t_k$ may be caused by a large current $i_k$
rather than a small threshold $t_k$.  Towards the end of the breakdown
process, this is the typical case.  We may therefore split the breakdown
process into three regimes:$^5$ (1) the disorder regime, where the bonds
break because they are weak so that it is the threshold distribution which
governs the breakdown process, (2) the competition regime where the
current distribution is roughly as wide as the threshold distribution,
causing the breakdown to be a subtle cooperative process between the
two distributions, and finally (3) the current-governed regime where
bonds break because they carry a large current.  This regime manifests
itself through a single macroscopic unstable crack eating its way through
the network, and eventually breaking it apart.  The disorder regime (1)
is characterized by nucleation of microcracks, and is essentially a
process in which bonds are cut at random (since there are no spatial
correlations in the way the thresholds were assigned).  The competition
regime (2) resembles superficially the disorder regime, but the long
range correlations that are developing through the current distribution
results in subtle scaling laws for example in the current-voltage
characteristics of the network.$^5$

The picture we have presented above is generic.  Nothing has been said
about the threshold distribution, $P(t)$.  It has earlier been argued that
in the limit of infinitely large lattices, the breakdown process is
completely determined by the strength of the singularities of the threshold
distribution for $t \to 0$ and $t \to \infty$.$^3$ These two singularities,
$P(t) \sim t^{\phi_0}$ for $t \to 0$ and $1-P(t) \sim t^{-\phi_\infty}$ for
$t \to \infty$, are characterized by the two exponents $\phi_0$ and
$\phi_\infty$.  Depending on these two exponents --- control parameters ---
the fracture process develops differently, even though the general
characteristics sketched above remains the same.  The behaviour of the
breakdown process may be classified into distinct {\it phases.\/} There are at
least five such phases.  (1) If either $\phi_0$ or $\phi_\infty$ is zero,
the disorder is so large that the current distribution is never able to
compete with it. The breakdown process in this case remains a random
percolation process until the lattice is broken apart. This is because the
only constraint on the breakdown process from the current distribution is
that the bond that may break carry a current different from zero.  This
leads to the breakdown process being a {\it screened percolation\/}
process,$^6$ and therefore in the universality class of standard
percolation.  In particular, a finite percentage of bonds must be broken in
order to break the network apart in the limit of infinitely large lattices.
(2) If both $\phi_0$ and $\phi_\infty$ are small --- a mean-field
calculation$^3$ puts both the critical $\phi_0$ and $\phi_\infty$ at the
value two --- the process is no longer in the universality class of
percolation, but still a finite percentage of bonds must be broken in order
to break the network apart.  (3) When $\phi_0$ is small (less than two
according to mean-field theory) and $\phi_\infty$ is large (i.e. larger
than two), the network is very weak in
the sense that only a fraction of bonds approaching zero needs to be broken in
order to break the network apart. More precisely, if the lattice has size
$L \times L$, the number of bonds to break scales as $L^{1.7}$ irrespective
of the two control parameters $\phi_0$ and $\phi_\infty$.  The fracture
process proceeds by a significant number of ``microcracks" --- small
clusters of broken bonds --- developing before one goes unstable and a
macroscopic crack eventually develops breaking the network apart.  (4) When
$\phi_0$ is large (larger than two according to mean-field theory) and
$\phi_\infty$ is small (smaller than two),
the number of bonds that breaks before the entire network ruptures
scales with the lattice size $L$ to a power smaller than two, but which,
unlike the previous case, now
depends on the two control parameters.$^7$ As with phase (3), the number of
bonds belonging to the final unstable crack breaking the network apart
forms a zero subset of the total number of bonds that break.  The key
difference between phases (3) and (4) in terms of the way the disorder
influences the fracture process, is that in phase (3) microcracks are being
induced because bonds are very weak, while in phase (4) new microcracks
open as those already in existence are stopped by very strong bonds. (5)
This is where both $\phi_0$ and $\phi_\infty$ are large.  According to a
mean-field calculation, ``large" means here larger than two.  An indirect
numerical calculation$^4$ puts the transition for $\phi_0 = \infty$ at
$\phi_\infty = 0.16$.  Another numerical calculation$^8$ puts for
$\phi_\infty \to \infty$, the critical $\phi_0$ at 1.4.  The defining
characteristics of this phase is {\it localization.\/} That is, few
microcracks form before one of them becomes unstable and eventually cuts
the lattice apart.  By ``few" we mean that the fraction of broken bonds
that do not belong to the unstable crack that eventually breaks the network,
goes to zero as $L\to \infty$.

In this letter we introduce the concept of {\it sensitivity\/} in
connection with fracture, borrowing it from the study of cellular
automata where it is known as$^9$ {\it damage spreading,\/}  a name
which would be very misleading in connection with fracture, as it has
nothing to do with the already well-established concept of damage in
fracture.  Let us define the concept operationally in terms of the fuse
network.  We set up two {\it identical\/} networks --- identical in the
sense that each corresponding fuse in the two lattices has the same threshold
value assigned to it.  The breakdown of these two copies will of course
then evolve identically.  Let us now choose a bond in, say, lattice
$A$ and set its threshold value to infinity, thus making it unbreakable.
The threshold of the corresponding fuse in lattice $B$ is set to zero, thus
making sure it will break immediately.  In this way we introduce a
small difference between the two copies, and the question we pose is: Will
the difference between the two lattices, in terms of where the cracks appear,
grow as the fracture process proceeds, or will stay small, that is of the
order of one bond.  If the difference grows, the networks are sensitive.

The interest in defining such a concept lies in the concept discussed in
the introduction, {\it tolerance.\/} In terms of the fuse network, we
imagine producing a set of such networks, all with the same distribution of
fuse strengths except for an added noise making each pair of corresponding
fuses slightly different.  The strength and distribution of this added
noise correspond to the tolerance.  If the added noise is sufficient to
induce the initial microcracks appearing under load to happen at different
places from copy to copy, and they are sensitive in the sense introduced
above, then the fracturing of different copies will develop differently.
In other words, predictions on how the other lattices will behave cannot be
made from testing one single copy.

Thus, the question of whether an added noise in the assignment of
thresholds is enough to change the breakdown properties of the network is a
question of the noise being strong enough to change where the initial
microcracks develop, and then if yes, whether the network is sensitive or
not.

Whether the noise is sufficient to change the initial microcracks is a
question of$^{10}$ {\it order statistics.\/} Let us assign the fuse
strengths according to the rule
$$
t_i=r_i^D\;,\eqno(1)
$$
where $r_i$ is a random number drawn from a uniform distribution between
zero and one.  $D$ is the control parameter; small $|D|$ values correspond
to small disorder and large $|D|$ values to large disorder.  In terms of
the cumulative threshold distribution, $P(t)$, which is the probability
to find a threshold value smaller than or equal to $t$, this corresponds to
$$
P(t)=\cases{t^{1/D}&    where $0<t<1$       if $D>0$;\cr
            1-t^{1/D}& where $1<t<\infty$  if $D<0$.\cr}
\eqno(2)
$$
Thus, in terms of the two control parameters $\phi_0$ and $\phi_\infty$,
we see that when $D<0$, $D=-1/\phi_\infty$, while $\phi_0 \to \infty$,
while for $D>0$, $D=1/\phi_0$, while $\phi_\infty \to \infty$.  We will
base our arguments on this distribution, even though it does not cover
all relevant disorders (for which both $\phi_0$ and $\phi_\infty$
simultaneously are finite). However, it is easy, as we will show, to
extrapolate our results into other regions of parameter
space.  Let us also assume that the cumulative
distribution of added noise is of the form
$$
R(\delta t)=\left({{\delta t}\over{\delta t_{\max}}}\right)^\eta\;,
\eqno(3)
$$
where $0<\delta t< \delta t_{\max}$, and $\eta >0$.

Let us now assume that $D>0$.  Then the threshold distribution for the bonds
{\it including\/} the noise is
$$
P_R(t)=\int_0^t d\tau\ \int_0^1 du\ \int_0^{\delta t_{\max}} dv p(u) r(v)
\delta\left( \tau-(u+v)\right) \;,
\eqno(4)
$$
where $p(t)=dP(t)/dt$ and $r(t)=dR(t)/dt$.  Integrating out the Dirac
delta function gives
$$
P_R(t)={\eta \over {D(\delta t_{\max})^{\eta}}}
\int_0^t du \int_{\max(u-\delta t_{\max},0)}^{\min(u,1)} dv\ v^{1/D-1}
(u-v)^{\eta-1}\;.
\eqno(5)
$$
For $t$ of the order of
$\delta t_{\max}$ or smaller, the distribution $P_R(t)$ behaves as
$$
P_R(t) = a t^{1/D+\eta}\;,
\eqno(6)
$$
where $a$ is a positive constant.  For larger $t$ it behaves as
$$
P_R(t)=P(t)=t^{1/D}\;.
\eqno(7)
$$
Suppose we draw $N$ ($=2\times L^2$) thresholds from the distribution $P(t)$.
We order them so that $t_{(1)} \le t_{(2)} \le ... \le t_{(N)}$.  The
expectation value for threshold number $k$ in this sequence is
$$
t_{(k)}=\left({k\over {N+1}}\right)^D\;,
\eqno(8)
$$
where we have used the general expression $P(t_{(k)})=k/(N+1)$.
We also form an ordered sequence of the thresholds obtained with the
perturbed distribution (5), $t'_{(1)} \le t'_{(2)} \le ... \le t'_{(N)}$.
For small values of $k$, the expectation value of the $k$th element
of this sequence is
$$
t'_{(k)}=\left({k\over {a(N+1)}}\right)^{D/(1+D\eta)}\;.
\eqno(9)
$$
We may now pose the question whether the added noise changes the sequence
of weak bonds or not?  If the sequence is changed, we have
$$
t'_{(k)} > t_{(k+1)}\;.
\eqno(10)
$$
Using equations (8) and (9) in this inequality leads to the expression
$$
1>a^{1/D\eta}\
\left(1+{1\over k}\right)^{1+1/D\eta} \left(k\over{N+1}\right)\;,
\eqno(11)
$$
In particular, for large $N$ {\it and\/} $k$, (11) may
be written as
$$
k < \left({1\over a}\right)^{1/D\eta}\ N\;,
\eqno(12)
$$
For any fixed $k$, (11) and (12) are always true for large enough $N$. Thus,
no matter how small the added noise is, it {\it does,\/} change the sequence
of the weakest bonds.  It should be noted in this argument that the upper
cutoff in the noise distribution, $\delta t_{\max}$, does not enter in the
discussion:  No matter how small it is, the noise will be relevant for the
weakest bonds when the network is large enough.

We now repeat this analysis for $D<0$.  The noise distribution is still
given by equation (3), while the threshold distribution now is
$$
P(t)=1-t^{1/D}\;,
\eqno(13)
$$
for $1<t<\infty$.  The threshold distribution after adding the noise is
$$
P_R(t)={\eta \over {D(\delta t_{\max})^{\eta}}}\
\int_1^t du \int_{\max(u-\delta t_{\max},1)}^u dv\ v^{1/D-1}
(u-v)^{\eta-1}\;.
\eqno(14)
$$
For $t$ close to 1, we have that
$$
P_R(t) = {a\over D} (t-1)^{1+\eta}\;,
\eqno(15)
$$
rather than
$$
P(t)={1\over D}\ (t-1)\;,
\eqno(16)
$$
for the unperturbed threshold distribution. Again ordering the sequence
of thresholds from the unperturbed distribution and the perturbed
distribution, equation (10) leads to the inequality
$$
1>(D^{D\eta}a)^{1/D\eta}\
\left(1+{1\over k}\right)^{1+1/D\eta} \left(k\over{N+1}\right)\;,
\eqno(17)
$$
which is satisfied for sufficiently large networks.
Thus, also in this case, the noise
will change the sequence of the fuses having the smallest thresholds.
As before, the upper cutoff of the added noise,
$\delta t_{\max}$ does not enter the discussion.

Thus, this chain of arguments leads to the conclusion that whether or not the
fuse network is sensitive to added noise does not depend on the noise
for large enough systems.  The next question is whether
it depends on the threshold distribution, $P(t)$, itself.
Thus, we investigate whether the system is {\it sensitive\/} or not
in the sense introduced above:  Starting with two identical copies of a
fuse network except for one pair of fuses which are made infinitely weak
and infinitely tough respectively, we measure whether the two copies develop
differently or not during breakdown.

It should be noted here that if we find that the network is sensitive with
respect to changing the threshold of only one bond, then it is sensitive
with respect to adding everywhere a noise to the threshold distribution.
However, the opposite is not true:  As we will see, in a certain regime
of disorder, $D>0$, the network is {\it not\/} sensitive with respect to
changing the threshold value of a single bond, but is sensitive with respect
to adding noise everywhere.

We have simulated the fuse network numerically generating ensembles
containing from 1000 to 200 samples each, and ranging in size from
$10\times10$ to $128\times 128$, using the threshold distribution (1), with
$-3< D < 1$.  Each time a fuse blows, we recalculate the current distribution
in what is left of the network using the conjugate gradient method.$^{11}$
This algorithm is eminently parallelizable, and ran very efficiently on a
Connection Machine CM5 computer.  Each sample consists of
two copies of the same network, but with one central bond different.  Both
networks are completely broken apart, and afterwards the macroscopic
crack breaking each of the two lattices is identified and compared.
The
order parameter we have used is
$$
S={{{1\over 2}\sum_{i,j} {\rm xor}\ (n^A_i,n^B_j)}\over
{\max\ (\sum_i n^A_i,\sum_j n^B_j)}}\;,
\eqno(18)
$$
where $n^A_i$ is one if bond $i$ of lattice $A$ belongs to the final
crack, otherwise it is zero.  $n^B_j$, likewise, concerns lattice $B$.
If the system is sensitive, we find that $S \to 1$, and if it is not,
$S \to 0$.  In figure 1, we plot $S$ as a function of the control parameter
$D$ defined in equation (1) for various lattice sizes.  As is evident,
there is a first order transition, {\it i.e.\/} a discontinuity, in the
order parameter $S$ for $D=0$, and a second order transition, {\it i.e.\/}
the slope of $S=S(D)$ diverges, for a negative $D=D_c$.  We determine
$D_c=-6.2\pm 1.0$ by plotting $D_{eff}(L)$ as a function of $L^{-1/\nu}$,
where $D_{eff}(L)$ is the $D$-value for lattice size $L$ where $S(D)$ has the
largest slope, and $1/\nu$ is chosen so that $D_{eff}(L)$ falls on a straight
line.  We show our fit in figure 2 and the exponent $1/\nu$ determined
>from here gives an estimate of the correlation length exponent $\nu=5\pm 2$.

Thus, we see that there is a window $-6.2 < D < 0$ in which the fuse network
is sensitive.  If $D>0$, then $\phi_0 =1/D$ and $\phi_\infty \to \infty$.
Within this range of $\phi_0$-values the network undergoes a
localization-delocalization transition, which numerical simulations$^8$
put at $\phi_0=1/D=1.4$.
There is no trace of this transition in the order parameter $S$.  For
negative $D$, there is a sensitive phase, which exists for
$\phi_\infty > 1/6.2=0.16$ and $\phi_0 \to \infty$.
The localization-delocalization transition in this range of parameters has
been, as already pointed out, numerically determined$^4$ to appear at
$\phi_\infty=-1/D=0.16$. The phase transition in $S$ seen at $D=-6.2$ is
therefore likely to be related to this transition.

Why does there seem to be a connection between the
localization-delocalization transition and the existence of a
sensitivity-insensitivity transition for $D<0$, but not for
$D>0$?  If, in the localized phase, the first bond
to break initiates the final macrocrack, we expect the localized phase to
coincide with the sensitive phase.  This seems to happen for $D<0$.
We may understand this by noting that in this case the
localization-delocalization transition is caused by crack
arrest:  The microcracks are {\it a priori\/} all unstable.  However, if the
distribution of strong elements is sufficiently large, the crack is stopped
by hitting a bond strong enough for the enhanced currents around the crack
tip to be insufficient to continue the growth of this particular crack.
Thus, when the disorder is small enough, the first microcrack cannot be
``held" back, and both sensitivity and localization follows.  On the other
hand, when the system is in the delocalized regime, a diverging (with the
lattice size) number of microcracks develop before one of them eventually
goes unstable and develops into a macroscopic crack.  Thus, a small initial
perturbation among the microcracks will typically not affect the
macrocrack that eventually develops, and the network is thus not
sensitive.

For $D>0$, the localized phase is different.  In this case
we have a localized phase even though there is a diverging number
(with lattice size) of
microcracks forming before one of them goes unstable and grows into the
macroscopic crack that breaks the network apart.  This is possible since the
ratio between the number of bonds forming the microcracks and the number of
bonds belonging to the final crack goes to zero: The total number of
bonds that has broken throughout the entire fracture process is dominated
by the final crack.  In this case, the
probability that the one artificially induced
microcrack in one of the two copies actually should be the one that
goes unstable falls to zero with the lattice size.
This happens since there is a power law distribution of
bonds whose thresholds are very weak, so that there always is a  ``mist"
of microcracks before one goes unstable, no matter how fast this
power law distribution falls off if the lattice is large enough.  Thus,
there will be no sensitive phase in this case, even though there is a
localized phase.

We now return to the question of sensitivity in connection with an
added noise in the threshold distribution.  In figure 3, we show
$S$ as a function of $D$ for an added disorder drawn from a
uniform distribution between zero and $\delta t_{\max}=0.1$.  There is
the same second order transition at $D=-6.2\pm 1.0$ in this case
as there is for the case when the difference between the two copies
is limited to one pair of bonds.  This is no surprise from the above
discussion. However, for $D>0$, there {\it is\/} a difference:  Now, there
is a sensitive phase for all $D>0$, while there was none when only a single
bond was changed.  We interpret this in the following way: As the added
noise affects the ordering of all of the weak bonds, and not only a single
one, we expect that also the one eventually leading to the final crack
is affected.

We conclude by recapitulating what has been found.  We have
investigated whether the fuse model is sensitive to the addition of
noise in the threshold distribution i.e. whether two
networks, identical except for the added noise, develop the same
macroscopic cracks or not.  The disorder in the fuse model is completely
described by two parameters.  We have investigated the sensitivity of the
model along a curve in this two-dimensional parameter space by two
very different types of noise. When a single bond is made unbreakable in
one copy and extremely weak in the other, we get a sensitive
region for $-6.2< D <0$. At the lower end the order parameter disappears
continuously with a large slope, while on the other side the order parameter
jump discontinuously to zero.  When a weak noise is added everywhere
the sharp jump at $D=0$ from a sensitive to an insensitive region
disappears and a sensitive region develops for $D>0$.
The negative $D$ region remains unchanged.
We identify this transition with a delocalization-delocalization transition.
The disappearance of the second insensitive region when going from strong
local disorder to weak nonlocal disorder is due to differences in crack
arrest mechanisms in the two parts of parameter space.

We thank D.\ Bideau, S.\ Roux and S.-z.\ Zhang for discussions in addition
to the unknown referee whose comments helped us improve the manscript
considerably.  We
also thank the IPG, the CCVR and the CIRCE (CNRS) for time on their
computers where these calculations were performed.  G.\ G.\ B.\ thanks
the University of Rennes for support during his stay at Rennes.  A.\ H.\
and G.\ H.\ R.\ thank the GRECO ``G{\'e}omat{\'e}riaux" and the GdR
``Milieux H{\'e}t{\'e}rog{\`e}nes Complexes" for financial support.
\vfill
\eject
\centerline{\bf References}
\bigskip
\frenchspacing
\item{[1]} L. de Arcangelis, S. Redner and H. J. Hermann,
J. Physique (Lett.) {\bf 46}, L585 (1985).

\item{[2]} {\sl Statistical Models for the Fracture of Disordered
Media,\/} edited by H. J. Herrmann and S. Roux (Elsevier, Amsterdam,
1990).

\item{[3]} A. Hansen, E. L. Hinrichsen and S. Roux, Phys. Rev. B
{\bf 43}, 665 (1991).

\item{[4]} A. Hansen, E. L. Hinrichsen, S. Roux, H. J. Herrmann and
L. de Arcangelis, Europhys. Lett. {\bf 13}, 341 (1990).

\item{[5]} L. de Arcangelis, A. Hansen, H. J. Herrmann and S. Roux,
Phys. Rev. B {\bf 40}, 877 (1989).

\item{[6]} S. Roux, A. Hansen, H. J. Herrmann and E. Guyon, J. Stat.
Phys. {\bf 52}, 237 (1988).

\item{[7]} S.-z. Zhang and A. Hansen, preprint.

\item{[8]} S.-z. Zhang, private communication.

\item{[9]} D. Stauffer, Phil. Mag. B {\bf 56}, 901 (1987).

\item{[10]} E. J. Gumbel, {\sl Statistics of Extremes\/} (Columbia
University Press, New York, 1958).

\item{[11]} G. G. Batrouni and A. Hansen, J. Stat. Phys. {\bf 52}, 747
(1988).
\vfill
\eject
\bigskip
\bigskip
\centerline{\bf Figure Captions}
\bigskip
\item{Fig. 1} The order parameter $S$, defined in equation (18), as a
function of the control parameter $D$, defined in equation (1) for
lattice sizes 16 (200 samples), 32 (200), 64 (200), and 128
(200).  The difference between the two copies constituting each sample is
the strength of a single bond.

\item{Fig. 2} $D_{eff}$ determined from figure 1 and additional lattice
sizes not shown plotted against
$L^{-1/\nu}$.  From this plot we estimate that $D_c=-6.2\pm 1.0$ and
$\nu=5\pm 2$.

\item{Fig. 3} The order parameter $S$ for 200 samples
of size $32\times 32$ and $64\times 64$ where the difference between
the two copies in each sample is a noise drawn from a
flat distribution between zero and 0.1 added to each bond (b).  We show
for comparison the corresponding curve for lattices of size
$64 \times 64$ when the difference between each copy is a single bond
((a), as in figure 1).
\bigskip
\vfill
\bye